\documentclass[aps,pra,twocolumn,nofootinbib,floatfix,10pt]{revtex4-2}

\usepackage{amsmath}
\usepackage{amssymb}
\usepackage{wasysym}
\usepackage{graphicx}
\usepackage{color,soul}
\usepackage{physics}
\usepackage{siunitx}
\usepackage{dsfont}
\usepackage{float}
\usepackage{ulem}
\usepackage{upgreek}
\usepackage{mathdots} 
\usepackage[english]{babel}
\usepackage{blindtext}
\usepackage[english,nomargin,inline,marginclue,draft]{fixme}
\pdfpageheight\paperheight
\pdfpagewidth\paperwidth

\usepackage[colorlinks,linkcolor=blue,anchorcolor=blue,citecolor=blue,urlcolor=blue]{hyperref}

\fxusetheme{colorsig}
\FXRegisterAuthor{cg}{acg}{CG}  
\FXRegisterAuthor{th}{ath}{\color{blue}TH}  
\FXRegisterAuthor{ib}{aib}{\color{red}IB} 
\FXRegisterAuthor{sh}{ash}{\color{cyan}SH} 
\FXRegisterAuthor{db}{adb}{\color{green}DB} 
\FXRegisterAuthor{ps}{aps}{PS}
\makeatletter
\renewcommand*\FXLayoutInline[3]{%
  {\@fxuseface{inline}\ignorespaces{\color{fx#1}[#3: #2]}}}
\makeatother

\long\def\symbolfootnote[#1]#2{\begingroup%
\def\thefootnote{\fnsymbol{footnote}}\footnotetext[#1]{#2}\endgroup}

\def\nobreakbefore{%
  \relax\ifvmode\else
    \ifhmode
      \ifdim\lastskip > 0pt\relax
        \unskip\nobreakspace
      \else 
        \nobreakspace
      \fi
    \fi
  \fi
}
\let\oldcite\cite
\renewcommand\cite{\nobreakbefore\oldcite}

\begin{document}
\title{Ultra-Wide Dual-band Rydberg Atomic Receiver Based on Space Division Multiplexing RF-Chip Modules}


\author{Li-Hua Zhang$^{1,2}$}
\author{Bang Liu$^{1,2}$}
\author{Zong-Kai Liu$^{1,2}$}
\author{Zheng-Yuan Zhang$^{1,2}$}
\author{Shi-Yao Shao$^{1,2}$}
\author{Qi-Feng Wang$^{1,2}$}
\author{Ma Yu$^{1,2}$}
\author{Tian-Yu Han$^{1,2}$}
\author{Guang-Can Guo$^{1,2}$}
\author{Dong-Sheng Ding$^{1,2,\textcolor{blue}{\dag}}$}
\author{Bao-Sen Shi$^{1,2}$}

\affiliation{$^1$Key Laboratory of Quantum Information, University of Science and Technology of China, Hefei, Anhui 230026, China.}
\affiliation{$^2$Synergetic Innovation Center of Quantum Information and Quantum Physics, University of Science and Technology of China, Hefei, Anhui 230026, China.}
\address{$^*$e-mail: dds@ustc.edu.cn}




\begin{abstract}
Detecting microwave signals over a wide frequency range has numerous advantages as it enables simultaneous transmission of a large amount of information and access to more spectrum resources. This capability is crucial for applications such as microwave communication, remote sensing, and radar. However, conventional microwave receiving systems are limited by amplifiers and band-pass filters that can only operate efficiently in a specific frequency range. Typically, these systems can only process signals within a three-fold frequency range, which limits the data transfer bandwidth of the microwave communication systems. Developing novel atom-integrated microwave sensors, for example, radio frequency (RF)-chip coupled Rydberg atomic receiver, provides opportunities for a large working bandwidth of microwave sensing at the atomic level. Here, an ultra-wide dual-band RF sensing scheme is demonstrated by space-division multiplexing two RF-chip-integrated atomic receiver modules. The system can simultaneously receive dual-band microwave signals that span a frequency range exceeding 6 octaves (300\,MHz and 24\,GHz). This work paves the way for multi-band microwave reception applications within an ultra-wide range by RF-chip-integrated Rydberg atomic sensor.
\end{abstract}
\maketitle

\section{Introduction}

Microwave electric field sensing has a wide range of applications, including cosmology detection \cite{spitler2016repeating}, remote sensing \cite{wooster2021satellite}, and communication \cite{du2010wireless}. Broadband Microwave reception technology enhances data transfer rates in communication \cite{8387217}, and improves resolution in imaging and remote sensing \cite{van2010high}. Due to the performance limitations of amplifiers and band-pass filters, typically, conventional receiver systems are designed for a specific frequency range and are unable to effectively handle signals with large frequency spans simultaneously. The upper-frequency limit of conventional receivers is generally within three octaves. In conventional RF sensing systems, most bandwidth of the amplifiers fails to exceed an octave with high efficiency over the entire operating range due to the limitation of the impendence match and transmission loss. To get the working bandwidth across octaves, engineers have to set the band edge frequencies to satisfy some conditions and constantly adjust the impedance of the system according to the frequency of the microwave in the sacrifice of efficiency \cite{wang2014,xuan2019,zheng2018,you2008}. In contrast, Rydberg atomic antennas have specific advantages in radio frequency (RF) detection, such as high sensitivity \cite{jing2020atomic,photonics9040250,ding2022enhanced}, self-calibration \cite{Selfcal2021}, low field perturbation, high accuracy \cite{sedlacek2012microwave}, a wide operating frequency range (from DC to terahertz), and non-destructive detection, making them free of thermal noise \cite{Meyer_2020,PhysRevApplied.18.014045,sayrin2011real, Waveguide2021,liu2022deep,10238372}. Besides, the frequency response of Rydberg atoms can be altered and extended by an extra RF field \cite{Hu2022continuous,PhysRevApplied.18.014033,PhysRevApplied.18.054003,PhysRevA.107.043102,PhysRevApplied.19.044049}.

\begin{figure*}[t]
\includegraphics[width=2\columnwidth]{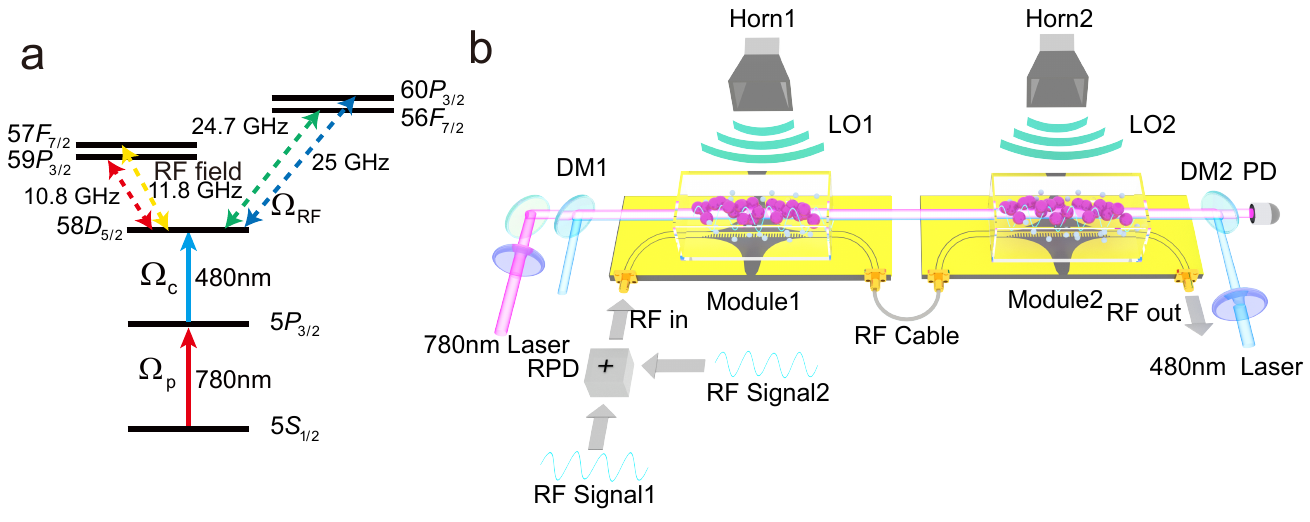}\caption{\textbf{Experimental energy level diagram and setup.} \textbf{a}, The energy level diagram of the Rb atom (Rubidium 85). The probe and coupling laser beams couple the atomic states of ground state $5S_{1/2}$, intermediate state $5P_{3/2}$, and Rydberg state $58D_{5/2}$ to realize electromagnetically induced transparency (EIT) configuration. The RF fields can couple the Rydberg state $58D_{5/2}$ with its adjacent Rydberg states $59P_{3/2}$, $57F_{7/2}$, $56F_{7/2}$ and $60P_{3/2}$ with different frequency intervals of 10.8\,GHz, 11.8\,GHz, 24.7\, GHz, and 25.0\,GHz (Dashed arrows). The coupling between the RF fields from 300\,MHz to 25\,GHz and Rydberg atoms can be separated into off-resonant and resonant regimes. For the RF fields with frequencies far from resonance, a strong local oscillator (LO) RF field is required to increase the system response, in which the AC-stark shifts induced by several pairs of Rydberg states off-resonant coupling. Near the resonant regime, the RF fields at frequencies 10.8\,GHz, 11.8\,GHz, 24.7\,GHz, and 25\,GHz drive Rydberg states with Autler-Townes (AT) effect. \textbf{b}, The multi-band microwave fields reception experiment setup. The whole detection system is composed of two receiver modules; each module consists of an RF chip and a centimeter-sized glass cell. A 780-nm probe laser and a 480-nm coupling laser are input into the two cells together with a dichroic mirror (DM). Then the EIT transmission signal of the probe laser is collected by a photodetector (PD). The information of RF signals (RF Signal1 and RF Signal2) can be read from the EIT spectrum output from the PD. Rydberg atoms are responded by the RF fields radiated by the LO field from the microwave horn and the signal field on the spoof-surface plasmon polaritons (spoof-SPP) chip. This supports an arbitrary tunable dual-band detection by using a pair of space-division multiplexing RF-chip-integrated Rydberg atomic receivers. The reception frequencies can be altered by changing the frequencies of the two strong LO fields emitted from the two microwave horns. Two RF signals (RF Signal1 and RF Signal2) with different frequencies are combined with a resistance power divider (RPD) and then input to the chip port.}
\label{fig1}
\end{figure*}

Rydberg atoms possess unique physical properties such as long coherence time and large polarization rate \cite{Adams_2020}, which makes Rydberg atoms sensitive to a wide range of RF signals and strongly coupled to broadband RF signals. Rydberg atoms hold promise in the reception of more information in a broader spectrum resource and enhancing bandwidth. For example, developing multi-band atomic antennas can enhance the spectrum utilization efficiency, increase system capacity, and adapt to different application scenarios \cite{haider2013recent,al2019mimo,7005405}. There are many works in this direction, for example, multi-band amplitude modulation (AM) and frequency modulation (FM) microwave communication
\cite{anderson2021} in a centimeter-sized glass cell, multi-band microwave reception \cite{holloway2019,anderson2021} in a glass cell with two different atomic species of Rubidium and Cesium atoms. By utilizing Rydberg RF transitions with different final states, multi-band reception and communication with inherently different transition frequencies are demonstrated \cite{10.1063/5.0095780,app10041346,PhysRevApplied.19.014025}. 
The integration of the RF metawaveguide chip with the Rydberg sensing system unleashes the high sensitivity potential of Rydberg atoms and extends the continuous working bandwidth of the atomic receiver. Based on this RF-chip-integrated design, we demonstrate a space-division multiplexing and modular dual-band microwave reception by detecting RF signals in two atomic receiver modules. The working frequencies in our system can be arbitrarily chosen within the ultra-wideband range of 300 MHz $\sim$ 24 GHz, exhibiting the ability of dual-band microwave processing across six octaves. This space-division multiplexing RF detection method based on RF-chip-integrated atomic receiver modules enables the integration of other microwave components into the system, making the system simple, compact, and scalable.

\begin{figure*}[htp]
\includegraphics[width=2\columnwidth]{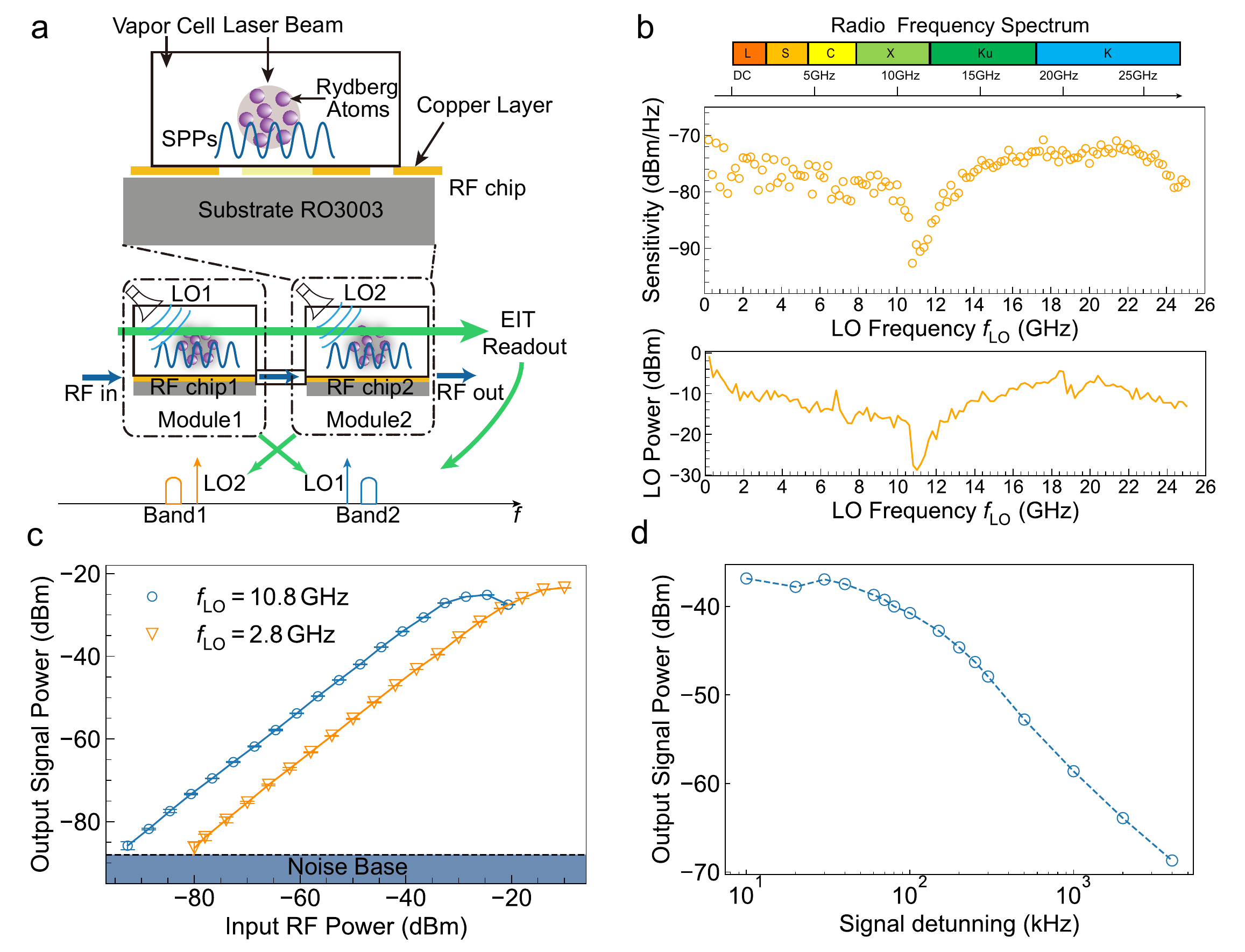}\caption{\textbf{The performance of the single RF-chip-integrated Rydberg atomic receiver module.} \textbf{a}, The structure of the RF-chip-integrated Rydberg atomic receiver. The integrated structure contains a spoof-surface plasmon polaritons (spoof-SPP) chip and a centimeter-sized glass cell. Over the chip surface, the RF signals in SPP mode are coupled with Rydberg atoms, and then the output signals are detected in an atomic heterodyne scheme by imposing LO fields with different frequencies. \textbf{b}, The intrinsic sensitivity for the RF-chip-integrated Rydberg system after a correction of the insert loss is plotted with orange circles. Each data points are acquired with a frequency separation of 200\,MHz. At 10.8\,GHz, the maximum sensitivity is measured at about -93\,dBm/Hz. In other frequencies (off-resonant region) above 300\,MHz, the sensitivity varies from about -70\,dBm/Hz to -80\,dBm/Hz. The detection range of the RF-chip-integrated Rydberg system covers the radio frequency band from Low Frequency to K-band (300\,MHz to 25\,GHz). The optimized LO field power applied for each frequency is plotted in Fig.~\ref{fig2}b, which varies from near 0\,dBm to -29\,dBm with the LO frequency. The power of the LO field has a similar variation tendency as the system sensitivities. \textbf{c}, The system dynamic range at 2.8\,GHz (blue circle) and 10.8\,GHz (orange triangles). The output signal varies in a similar range from about -25\,dBm to -86\,dBm (equal to the system noise base), the resonant sensitivity at 10.8\,GHz is about 13\,dB larger than the off-resonant situation at 2.8\,GHz. The error bars come from experimental statistics. \textbf{d}, The system instantaneous bandwidth. We scan the signal RF detuning $\delta_{\mathrm{sig}}$ from the LO field ($f_{\mathrm{LO}}$ = 2.8\,GHz), and set the power of the signal RF field to -25\,dBm (About -28\,dBm at the input port). The instantaneous bandwidth of the system is about 100\,kHz at a 3dB reduction.}
\label{fig2}
\end{figure*}

\section{Results and Discussion}

\textbf{Experimental Setup}
As the energy level diagram illustrated by Fig.~\ref{fig1}a, we excite the Rubidium atoms from ground state $5S_{1/2}$ to Rydberg state $58D_{5/2}$ through a two-photon electromagnetically induced transparency (EIT) scheme \cite{PhysRevLett.98.113003}. The atoms at the Rydberg state $58D_{5/2}$ can be driven by the input LO and signal RF fields with several transitions of $59P_{3/2} \leftrightarrow 58D_{5/2}$, $58D_{5/2} \leftrightarrow 57F_{7/2}$, $56F_{7/2} \leftrightarrow 58D_{5/2}$, and $58D_{5/2} \leftrightarrow 60P_{3/2}$. For simplicity, we only consider the transitions from $58D_{5/2}$ to the adjacent  ($\Delta n = \pm 1$)  and next-adjacent  ($\Delta n = \pm 2$) Rydberg state, as their dipole moments are larger compared with other dipole-allowed RF transitions. The frequencies of these transitions are 10.8 GHz, 11.8 GHz, 24.7 GHz, and 25.0 GHz, respectively. The resonant frequencies of these transitions that are calculated by the python package \cite{vsibalic2017arc}  cover a large range. The Rydberg atomic receiver has an enhanced working bandwidth due to both the resonant and off-resonant coupling with RF fields by the atomic RF transitions mentioned above. The effective coupling between RF fields and Rydberg atoms is realized via the (resonant) Autler-Townes (AT) effect and the (off-resonant) AC-stark effect.

\begin{figure*}[htpb]
\includegraphics[width=2\columnwidth]{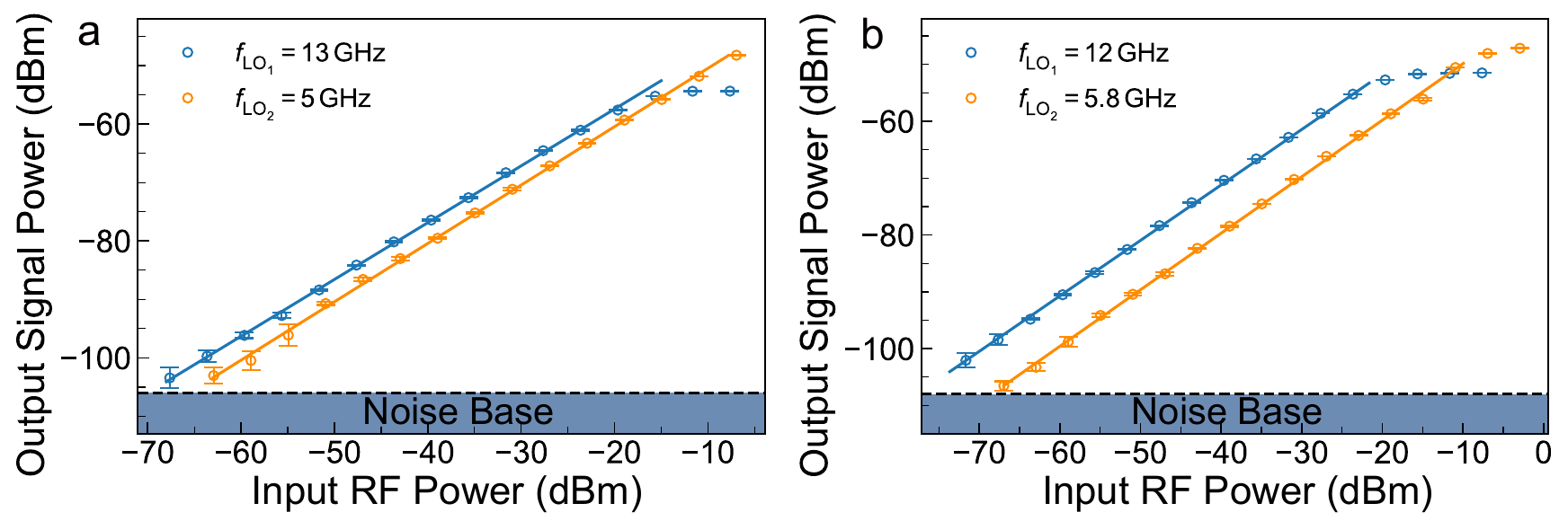}\caption
{\textbf{The dynamic range for dual-band microwave reception.} RF signals at arbitrarily selected frequencies (5\,GHz and 13\,GHz, 5.8\,GHz and 12\,GHz) are received and the corresponding dynamic ranges are plotted. \textbf{a}, The input RF signal with a frequency of $f_{\mathrm{sig}}$ = 5\,GHz + 20\,kHz, and $f_{\mathrm{sig}}$ = 13\,GHz + 20\,kHz, the linear dynamic ranges are about 57 dB at 5\,GHz and 55\,dB at 13\,GHz. \textbf{b}, The dynamic ranges are about 56\,dB at 5.8\,GHz and 52\,dB at 12\,GHz. The error bars come from experimental statistics.}
\label{fig3}
\end{figure*}

Fig.~\ref{fig1}b shows the experiment setup of the dual-band microwave reception based on RF-chip integrated Rydberg atomic receivers. The single reception module in the experiment consists of a vapor cell and an RF chip below the cell. The structure of a single module and the space-division multiplexing of dual-band detection schemes are given in Fig.~\ref{fig2}a. The spoof-SPP RF chip is utilized to enhance the electric field intensity over the region where Rydberg atoms are excited by the laser beams in a vapor cell with a 50-mm length. The coupling laser (Toptica, DLpro 480) and the probe laser (Toptica, DLpro 780) counter-propagate with each other over the periodical meta-structure on the chip, and the beams are focused in the centimeter-sized vapor cell. Based on the single reception module, dual-band reception is achieved by sharing the RF signal between the two single modules. Different frequency bands are selectively detected through different LO fields. The free-space LO field $E_{\mathrm{LO}}$ beat with the RF signal field $E_{\mathrm{sig}}$ over the chip and the output beat signal with a frequency of $\delta_{\mathrm{sig}} = |f_{\mathrm{LO}}-f_{\mathrm{sig}}|$ is read from the Rydberg EIT spectrum with a photodetector (Thorlabs, PDA36A). 

\textbf{Ultra-wide working bandwidth}
The dual-band microwave reception experiment requires arbitrarily tunable RF frequency working points. This is achieved by the off-resonant atomic heterodyne method \cite{Meyer_2020,Waveguide2021,PhysRevApplied.18.014045,Hu2022continuous}. For the operation of the Rydberg atomic receiver under off-resonant conditions, the response of Rydberg atoms to the off-resonant RF signal is improved by applying a strong LO field. The strong LO field couples many adjacent Rydberg states with the initial Rydberg state $58D_{5/2}$ and induces strong AC-stark energy shifts on the energy levels of Rydberg atoms.  The weak RF signal beats with the LO field by perturbing the total RF field sensed by the atoms. Finally, the amplitude and phase of the signal RF field are read from the beat signal of the Rydberg EIT spectrum.  In our experiment, the frequencies of the strong LO RF field and signal RF field were scanned from 300\,MHz to 25\,GHz, and the detuning of the signal RF field from the LO field $\delta_{\mathrm{sig}} = f_{\mathrm{sig}}-f_{\mathrm{LO}}$ was set to $20\,$kHz at each tested frequency point. The output beat signal amplitude for the EIT-AT spectrum was then recorded to calculate the intensity of the input weak signal RF field. To estimate the sensitivity of the system, we need to calibrate the insertion loss of coaxial cables and a resistance power divider, which is given in Fig.~\ref{fig1}b. The sensitivity is obtained based on the beat-note response in the linear region and is in inverse ratio to the photodetector noise.

To ensure the accuracy of the measurement, the input power of the RF signal field was set within the linear zone of the atomic heterodyne response. The sensitivity of the system was found to be lower below 300\,MHz compared to the range of 300\,MHz to 25\,GHz. This is attributed to manufacturing errors of the spoof-SPP chip as indicated by the S21 parameter, see the Method Section. At low RF frequencies, the propagation constant of the SPP surface wave reduces, leading to weak field confinement over the spoof-SPP structure. In addition, the screening effect of the vapor cell causes a decrease in the intensity of RF fields in the range of a few kHz \cite{below1khzJau}. While, at frequencies above 300\,MHz, the system shows obvious enhancement when the RF fields resonate with the Rydberg transitions. Due to the similar coupling strength of Rydberg atoms with the RF field (either at 10.8 GHz and 11.8 GHz or at 24.7 GHz and  25 GHz) and the fluctuation of the chip performance, it is hard to distinguish all the four resonant peaks for Rydberg atoms in Fig. \ref{fig2}b. Two resonant peaks are observed at approximately 10.8\,GHz and 25\,GHz. At 10.8\,GHz, the maximum (input) sensitivity of the system is around -93\,dBm/Hz. Although there is a resonant response at 25\,GHz due to the resonant transition, it is relatively weaker compared to the 10.8\,GHz peak due to the reduction of the transition strength and spoof-SPP chip loss. The system sensitivity of the single module below 11\,GHz has more fluctuations, this is partly due to the uneven conversion for electric field intensity of the chip at a lower frequency. The RF transitions of ($55F_{7/2}\leftrightarrow58D_{5/2}$ and $58D_{5/2} \leftrightarrow 61P_{3/2}$) cover the frequencies 60\,GHz (59.001\,GHz and 63.218\,GHz), these frequencies are far away from the frequency of the input RF field. Due to the larger detuning of the RF fields with transitions $55F_{7/2} \leftrightarrow 58D_{5/2}$ or $58D_{5/2} \leftrightarrow 61P_{3/2}$, the AC-stark shift for the Rydberg states $55F_{7/2}$ and $61P_{3/2}$ is small. In addition, the degradation of chip performance also contributes to a decrease in system sensitivity above 25 GHz. Utilizing lower-$n$ Rydberg energy levels to increase the off-resonant response range of Rydberg atoms and optimizing the metawaveguide structure may help to further improve the system's operating range.

\begin{figure*}[htpb]
\includegraphics[width=2.05\columnwidth]{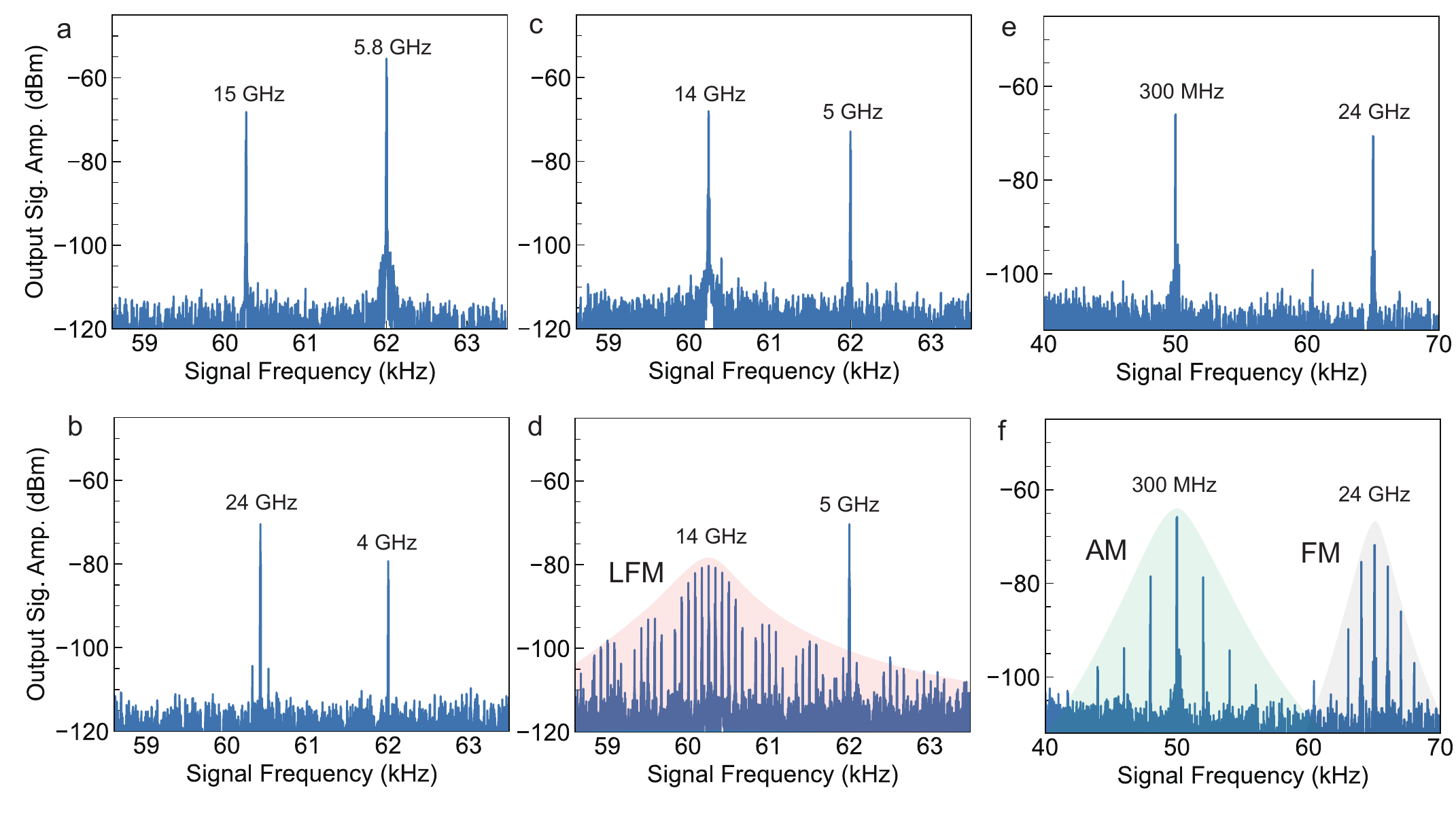}\caption
{\textbf{The arbitrary dual-band signal spectrum received simultaneously by two RF-chip-integrated atomic receiver modules.} \textbf{a-f}, The high-frequencies of the LO fields (applied to module 1) $f_{\mathrm{LO_{1}}}$ are set to 15\,GHz, 14\,GHz and 24\,GHz, and the low-frequencies of the LO fields (applied to module 2) $f_{\mathrm{LO_{2}}}$ are set to 5.8\,GHz, 5\,GHz, 4\,GHz and 300\,MHz, respectively. \textbf{c}, The frequency spectrum for the input signals with central frequencies of 14\,GHz and 5\,GHz. \textbf{d}, The frequency spectrum for the input signals with the same central frequencies, but with a linear frequency modulation applied on the 14\,GHz signal (LFM, marked by red shaded area). In these measurements, the RF signal with a frequency offset of $\delta_{\mathrm{sig}}$ = 60.2\,kHz and 62\,kHz from the corresponding LO signal are received by two atomic sensor modules. \textbf{e-f}, The frequency spectrum for simultaneously receiving at 0.3\,GHz and 24\,GHz, in which amplitude modulation (AM, marked by green shaded area) and frequency modulation (FM, marked by gray shaded area) are applied to the RF carrier with frequencies of 0.3\,GHz and 24\,GHz.(\textbf{f}). In these measurements, the RF signal with a frequency offset from the LO field $\delta_{\mathrm{sig}}$ is set as 50\,kHz and 65\,kHz, respectively. All of the spectra are sampled by an electric spectrum analyzer (Ceyear 4024F) with an RBW (and VBW) of 3\,Hz and a sweep time of 3.277s in sweep mode. }
\label{fig4}
\end{figure*}

However, the utilization of lower-$n$ Rydberg states may introduce weaker off-resonant and resonant sensitivity due to their smaller polarizability and dipole transition moment. Different optimal LO powers are applied for LO frequencies near resonance and far from resonance to maximize the output signal, as shown in Fig.~\ref{fig2}b. At 10.8\,GHz, the power is set to -29\,dBm. Furthermore, in the off-resonant heterodyne scheme ($\Delta_{\mathrm{RF}} \gg \Gamma_{\mathrm{EIT}}$), where the RF signal detuning is larger than the EIT bandwidth), a LO field much larger than that in the resonant scheme is required. Nevertheless, the field enhancement feature of the spoof-SPP chip helps to increase the coupling strength between the RF fields and Rydberg atoms thus alleviating the power requirement for the LO field, enabling optimal operation under off-resonant conditions. This in turn reduces the system's energy consumption, facilitating further miniaturization.

\textbf{The dynamic range and instant bandwidth of the single receiver.}
The dynamic range and instantaneous bandwidth of the RF-chip integrated Rydberg atomic receiver determines how weaker and how much signal can be received. The dynamic range of a dual-band receiver refers to the difference in signal strength that it can simultaneously handle, ranging from the maximum to the minimum. The total dynamic range of our system is about 70\,dB, and a linear dynamic of about 60\,dB is shown in Fig.~\ref{fig2}c at 1-Hz resolution bandwidth measured by the electric spectrum analyzer. The sensitivity and dynamic range are finally limited by a noise base of about -90\,dBm from the optical read-out. In the measurement of dynamic range, the LO and signal RF fields are first combined by a resistance power divider, and then are fed into the spoof-SPP chip. The signal dynamic range for the RF signals in different frequency bands is individually tested but not input together. The dynamic achieved here is not the cutting-edge which is mainly due to the low coupling laser power.
By increasing the Rydberg excitation efficiency with a stronger coupling laser system to increase the EIT signal intensity or locking lasers to a reference ultra-stable cavity to reduce the read-out noise \cite{jing2020atomic}, the dynamic range of our system can be improved further. 

The instantaneous bandwidth of a dual-band receiver refers to the frequency range that the receiver can simultaneously process at a given moment, referring to the simultaneous signal frequency reception range and processing capability of the receiver. By inputting the signal RF field with detuning $\delta_{\mathrm{sig}}$ from the LO RF field ($f_{\mathrm{LO}}$ = 2.8\,GHz), we can measure the instantaneous bandwidth of the system. The results are given in Fig.~\ref{fig2}d, in which an instantaneous bandwidth of about 100\,kHz is obtained. The bandwidth is ultimately limited by the Rydberg-EIT bandwidth, which is near 10\,MHz. Thus, the bandwidth for our system can be further optimized by reducing the beam waists (focusing the laser beams) or adjusting the coupling and probe laser powers \cite{PhysRevApplied.18.034030}. Although a relatively small instantaneous bandwidth of about 100\,kHz is obtained in our experiment, there are also some potential applications in which the reception of narrowband signals is required, such as radio frequency identification (RFID) and navigation systems. In these cases, the receivers work with smaller instantaneous bandwidth. The instantaneous bandwidth could exceed 10\,MHz working in a six-wave mixing scheme \cite{yang2023highsensitive}.

\textbf{Dual-band atomic microwave reception based on space-division multiplexing}. 
Distinct from conventional receivers based on absorption detection, the EIT-based Rydberg atomic RF sensor can detect the RF field with a little perturbation to the input signal \cite{Meyer_2020, Waveguide2021}, thus a dual-band multi-channel detection can be achieved by space-division multiplexing two chip-integrated modules. After the characterization of the ultra-wide working bandwidth of the RF-chip integrated Rydberg atomic receiver, we perform RF signal reception with tunable dual-band frequencies by using two modules (module 1 and module 2 as shown in Fig.~\ref{fig1}b and Fig.~\ref{fig2}a. These two modules are spatially separated but connected with a low-loss RF cable. Heterodyne detection is also applied in this case, but here we apply the LO RF fields in free space by a pair of horn antennas to reduce the crosstalk between the LO fields in a space-division multiplexing scheme. (In principle, the LO RF fields can be fed through the chip ports directly, but the energy shift of the two strong LO fields would couple with Rydberg atoms in the same vapor cell and influence the optimal working point for each module). Due to the low perturbation feature of the Rydberg atomic receiver, module 1 rarely affects the RF signal detected by module 2. Module 1 only causes a small reduction of the sensitivity for module 2 due to the small transmission loss (which is mainly determined by the transmission coefficient of the single RF chip). Based on this advantage, we can achieve a tunable dual-band RF detection by only monitoring the probe beam after module 2. The minimum frequency separation that the two modules can work without affecting each other is of the same order of magnitude with the EIT bandwidth (about 10 MHz). As the system is operated mainly under off-resonant conditions, the dispersion and absorption of the RF signal by the previous module (module 1) is small. However, the leakage of a strong off-resonant LO field between the modules is measurable. Therefore, the separation of the operation frequencies between the two modules needs to be much larger than the system’s instantaneous bandwidth (100 kHz) to minimize the crosstalk between the modules. To reduce the insert loss of the system to receive the dual-band RF fields, module 1 is applied to detect the signal above 10\, GHz, and module 2 is used to receive the signal below 10\,GHz. The spoof-SPP chip has a less loss below 10\,GHz, for example, the total insert loss measured for the signal at 5.8\,GHz (containing the loss of the first chip and the RF cable) is about 1.65\,dB. In Fig.~\ref{fig3}a, the dynamic range and the output signal power of the dual-band RF-chip-integrated Rydberg atomic receivers are plotted. The frequency of the LO field of module 1 $f_{\mathrm{LO_{1}}}$ is set to 13 GHz and the frequency of module 2 $f_{\mathrm{LO_{2}}}$  is set to 5 GHz in Fig.~\ref{fig3}a to receive RF signals with an offset of 20\,kHz from LO fields. In Fig.~\ref{fig3}b, $f_{\mathrm{LO_{1}}}$ is set to 13 GHz and $f_{\mathrm{LO_{2}}}$ is set to 5.8 GHz.

In our experiment, there is a sensitivity reduction of about 10 dB compared with the single-module scheme that the LO, and signal RF fields are put together through the same input port, this is from the mismatches between the SPP fields over the chip and the free space field from the horn antenna. The mode of the LO field emitted from the rectangular waveguide antenna is the TE (Transverse Electric) mode. Whereas the SPP is the TM (Transverse Magnetic) mode \cite{maier2007plasmonics}, so RF field propagating on the chip does not match the TE mode in the interaction with Rydberg atoms. This mismatch can be corrected by further technical upgrades such as adding more chips to replace the microwave horn.

The power of the LO RF fields for 5.8\,GHz and 5\,GHz being input to the horn antenna is about 4\,dBm. There is about 15\,dB larger than the power of the LO RF field which is directly input through the spoof-SPP chip port. Considering the strong LO field applied to the receiver module, the multi-path diffraction and reflection of the LO field will increase the phase noise of the output signal and reduce the sensitivity of the system \cite{1570023}. Integrating the wave-absorption material into the reception module and optimizing the metal structure of the chip may help to reduce the diffraction and reflection effect. Also, this effect will be suppressed if the LO field is applied by another chip efficiently from another side of the vapor cell. As shown in Fig.~\ref{fig4}, the output frequency spectrum in different arbitrary frequencies dual-band receiving by module 1 and module 2 are plotted. We record this spectrum by setting the RF signal detuning from corresponding LO fields $\delta_{\mathrm{sig}}$ to 60.2 kHz (module 1)  and 62 kHz (module 2), respectively.  The dual-band signal reception at three arbitrarily chosen dual frequencies is demonstrated. The frequencies being chosen are 15\,GHz and 5.8\,GHz (Fig.~\ref{fig4}a), 24\,GHz and 4\,GHz (Fig.~\ref{fig4}b), 14\,GHz and 5\,GHz (Fig.~\ref{fig4}c). The results show the system's ability to receive dual-band signals simultaneously. The input signal power before module 1 is about -35\,dBm, -30\,dBm and -20\,dBm for 4\,GHz, 5\,GHz, and 5.8\,GHz, respectively. All the power of these signals is measured by a frequency analyzer (Ceyear, 4024F).

In addition, dual-band simultaneous reception with different modulations for each RF signal is also demonstrated. Here, we set different frequency modulations for the two modules by applying a linear frequency modulation for module 1 (with a center frequency of 14\,GHz) and applying a single frequency (5\,GHz) for module 2. The results are given in Fig.~\ref{fig4}d, many modulation sidebands are observed around 14\,GHz, while only a single narrow peak appears at 5\,GHz. This result is in good agreement with the modulation. We also demonstrate RF receiving with a large range of 300\,MHz and 24\,GHz, the results are shown in Fig.~\ref{fig4}e and Fig.~\ref{fig4}f. As shown in Fig. \ref{fig4}f,  the RF signals with different modulations and different central frequencies are received simultaneously (300 MHz and 24 GHz). Thus, the spectrum information in two bands is captured, which means a faster data reception rate is achieved compared with the single reception module. For a single receiver module, only the RF signal in a certain band (either near 300 MHz or 24 GHz) is received. Here, the RF signal at 300 MHz is amplitude modulated (AM) with a modulation frequency of 2 kHz and the RF signal at 24 GHz is frequency modulated (FM) with a modulation frequency of 1 kHz.  This dual-band RF reception cross six octaves shows the multi-band information processing ability of the RF-chip-integrated Rydberg atomic device. 

\section{Discussion}
The power sensitivity of the chip system is mainly limited by the electric field response and the coupling strength between Rydberg atoms and the RF chip. As Rydberg atoms have a direct response to the electric field strength and are excited through a non-absorptive detection process, it is necessary to convert the power sensitivity of an RF signal to the intrinsic electric sensitivity of Rydberg atoms \cite{9374680}. A field enhancement scheme can be employed to improve the coupling strength between the atoms and the different RF signals, which improves the overall performance compared to a simple free-space sensing scheme \cite{PhysRevA.105.022626,yang2023local}. In addition, the multi-band RF field enhancement devices incorporating metal components like waveguides or resonators may introduce additional thermal noise or bandwidth limitations to the Rydberg sensing system \cite{PhysRev.32.110,1126767}. In our scheme, Rydberg atoms in different modules participate in the reception of fields at different frequencies, thus reducing direct crosstalk between Rydberg atoms.

By integrating the RF field, Rydberg atoms, and optical field on a chip, the sensing system can be made more portable, affordable, and accessible for various applications. The chip sensor in this work without larger vacuum chambers exhibits a more simplified sensor. This multi-band integrated scheme can incorporate various functional modules on the RF chip to realize multiple sensing applications. This integration also allows for compact and lightweight designs, reduces the power requirement for a strong LO field in Rydberg heterodyne sensors benefiting from the field enhancement setup, and minimizes overall costs. One can integrate a large number of RF-chip-integrated Rydberg atomic receivers to demonstrate a sensor array in which each integrated Rydberg atomic receiver could implement a distinct function. Through this method, multiple Rydberg atoms can participate in the measurement compared with the single module scheme providing extra gain of the sensitivity. Besides, the sensing module in the system is easily extensible due to the simple setup and the signal for each module can be read out individually and conveniently if fiber-coupled cells are used. In this case, a faster data reception rate compared with the single atomic receiver can be achieved. Notice that a low-noise RF amplifier before the input port may help to largely improve the sensitivity \cite{Waveguide2021}, however, extra thermal noises would finally limit the system performance to the thermal noise limitation (-174\,dBm/Hz). With the quick improvements in the sensitivity of the Rydberg atomic electrometry \cite{tu2023approaching} and the combination of the field-enhance scheme, the thermal noise limitation for the conventional receiver can be beaten \cite{ieee2021commu,santamariabotello2022comparison}.

\section{Conclusion}
In this study, we research multi-band microwave reception using RF-chip-integrated Rydberg atomic receivers experimentally. These receivers can operate across a wide frequency range, from 300\,MHz to 24 GHz, enabling the measurement of microwave signals in multiple frequency bands simultaneously. The RF-chip-integrated Rydberg atomic receivers lie in their ability to receive and process multiple frequency bands simultaneously using a single device. A dual-band RF signal that covers a frequency range exceeding six octaves is received by the RF-chip-integrated Rydberg atomic receivers, in which the AM and FM modulation signals at the different frequency bands are successfully obtained. By integrating the functionalities of both the chip and atoms into a single device, the complexity of the multi-band atomic sensors is reduced, and there is potential for the development of compact and highly integrated systems.

Traditional RF-chip-integrated receivers are designed for specific frequency bands, requiring separate devices for each band, low-noise amplifiers, and complex filters that bring extra thermal noise and frequency restrictions. In contrast, multi-band microwave reception based on Rydberg atomic receivers can cover multiple bands with a single device, improving receiving efficiency.

This is particularly important in wireless communication and radar systems, where signals span multi-band bands, and simultaneous reception for multi-band signals allows for parallel processing of signals, enhancing overall system performance. The reported results show a new-type continuous tuning dual-band RF reception by the Rydberg atomic system is modularized and scalable which is promising for communications and metrology.
\section{Methods}
\textbf{Design and Characterization} 

\begin{figure*}
\includegraphics[width=2\columnwidth]{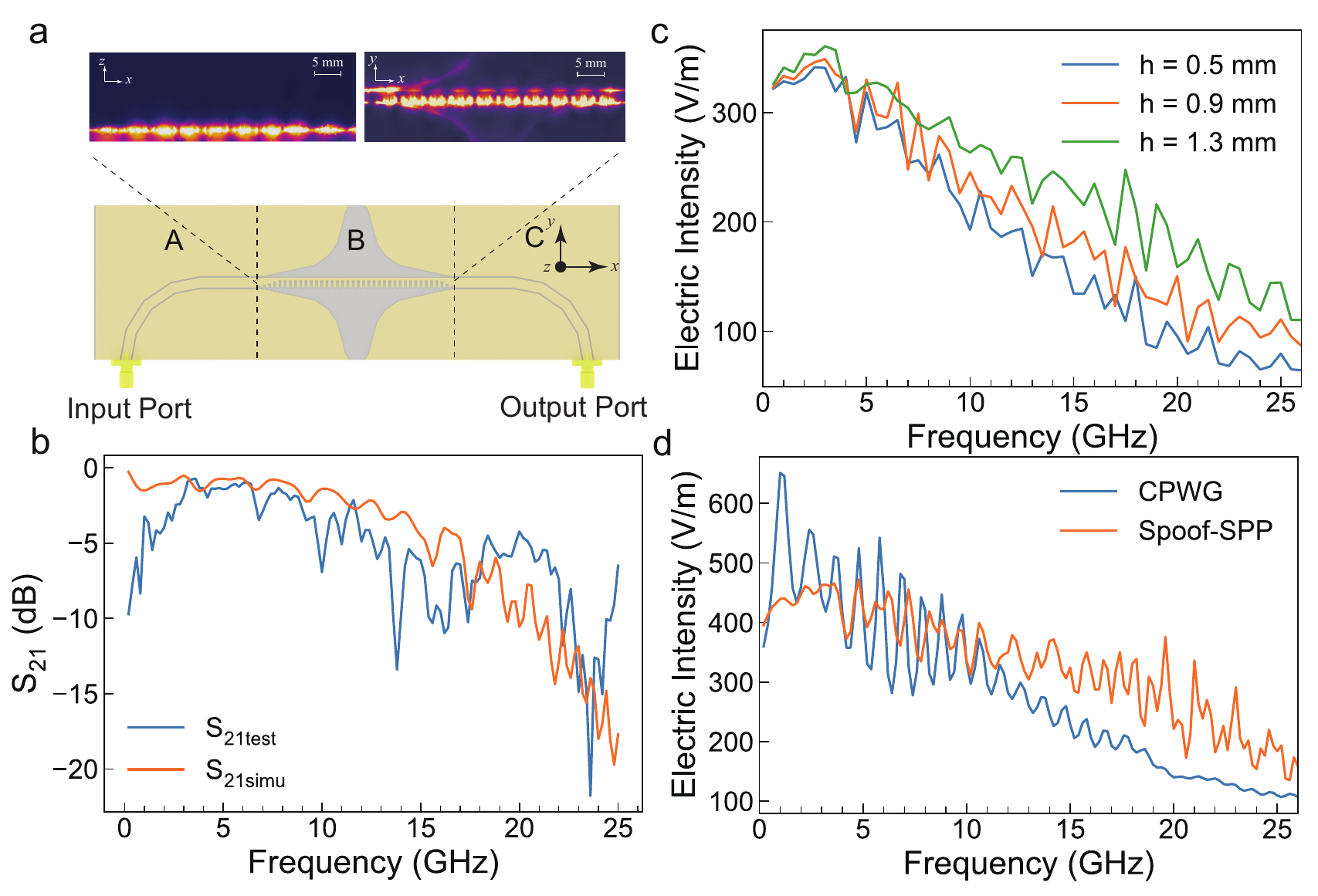}\caption
{\textbf{The properties of the spoof-SPP chip.} \textbf{a}, The structure of the spoof-SPP chip and the steady field intensity distribution in $zx$ and $xy$ surface over the spoof-SPP region with an RF signal with a frequency of 18\,GHz input. The RF field transmits through the region of A, B, and C on spoof-SPP chips with quasi-transverse electromagnetic (TEM) mode, SPP, and quasi-TEM modes.  \textbf{b}, The experimental tested and simulated results of S21 parameters for the chip. \textbf{c}, The simulated averaged electric field intensity over the chip surface for different groove depth $h$ of spoof-SPP structure are plotted. \textbf{d}, The simulated intensities of the electric field for the CPWG in Ref.~\cite{Waveguide2021} and Spoof-SPP in this article are illustrated. In these simulations, the power at the input port of the chip is set to be 20\,mW. The region where the electric field strength is averaged is cylindrical. The radius of this region is 0.3\,mm and the length is 50\,mm. }
\label{fig5}
\end{figure*}

As the spoof-SPP waveguide cannot be fed efficiently from the SMA port, a curved coplanar waveguide (CPW) line is introduced. As shown in Fig.~\ref{fig5}a, a gradient transition structure in region B is used to finish the mode conversion from the CPW line to the spoof-SPP line \cite{ssppconver}. The chip is plated with a 35-$\mu$m copper layer on the Rogers 3003 substrate. 

As the output signal of the system depends on the electric intensity of the RF signal sensed by Rydberg atoms \cite{sedlacek2012microwave,jing2020atomic,Waveguide2021}, the average field intensity over the chip is simulated by the finite element methods. Besides, the transmission coefficients (S21) for the chip are also simulated and experimentally tested, which is shown in Fig.~\ref{fig5}b.
The difference between the simulation and experimental results of the S21 parameter is mainly due to the fabrication error and the permittivity change of the vapor cell wall compared with the simulated air media. The dispersive property of the SPP wave can be tuned by the grove depth $h$, and a larger $h$ leads to a lower cut-off frequency. The grove depth $h$ is set to 1.3\,mm to balance the wide-band transmission and the constraint ability of the electric field. 

The simulation results of the average electric field intensity at a distance of 1 mm above the chip, in a cylindrical region with a radius $\mathrm{r}=300\,\mu m$, are depicted in Fig.~\ref{fig5}c. It is proved that a deeper groove depth helps to increase the field intensity over the chip. Further design optimization is necessary to minimize loss, enhance electric field concentration, and improve detection sensitivity. 

At the same time, the coupling loss between the chip and atoms needs to be taken into consideration. This loss is primarily composed of two factors. Firstly, there is a mismatch between the region where the atoms are shinned and the volume where the RF field is concentrated. Secondly, a portion of the SPP wave propagates in the substrate, resulting in the electric field not coupling with the atoms. As a result, a part of the effective RF power is lost when it is converted into electric strength on the chip surface.

\textbf{Simulation of the electric field conversion for the chip}

The electric field intensity over the chip is simulated by the finite element method. The average electric field over the chip in a cylinder with a radius of 300\,$\mu m$ for different groove depths is simulated and plotted. Here, our simulation is based on a steady-state solver so the time-average effect for the electric intensity is considered to be the space averaged due to the periodic distribution of the RF field. Furthermore, a time-dependent transient analysis may be more accurate about the averaged field intensity. 

The average field intensity against frequencies for both the coplanar waveguide (CPW) in Ref.~\cite{Waveguide2021} and the spoof-SPP chip in this article are simulated and compared. As shown in Fig.~\ref{fig5}d, the CPW waveguide has a more effective concentration below 5\,GHz because the SPP mode is not perfectly matched, but the variation against the frequency is greater compared with the spoof-SPP chip. The spoof-SPP chip has a slower degradation of the performance and has better field-enhancement features above the 10\,GHz compared with the CPW waveguide. and has a better field-enhancement performance above 10\,GHz. The simulated electric field intensity has a small reduction at low frequency (below 2\,GHz).  
 
Besides, the vapor cell causes a significant index change effect for the RF field over the chip, which needs to be taken into account. The Rydberg vapor in the cell has a small index change compared with other components in the reception module. However, there is an index change of the RF field from the glass wall of the vapor cell \cite{Meyer_2020}. This leads to a change in polarization and a reduction in the strength of the RF field. Using low-dielectric glass materials \cite{PhysRevApplied.13.054034}, embedding the chip in a larger vapor cell \cite{Waveguide2021}, or reducing the thickness of the glass wall are effective ways to suppress this effect. A finite element simulation including the vapor cell over the chip will reveal more details.

\textbf{Experiment Details} 

The probe beam with a power of 15 $\mu$W which is emitted from a 780-nm diode laser (Toptica, DL pro) has a Rabi frequency of $\Omega_{p}$= 6.62 MHz (for each probe beam). The probe laser is locked to the atomic transition $\mathrm{^{85}Rb}$ $F=3 \rightarrow F^{\prime}=4$. The coupling beam with a power of 80\,mW and  $\Omega_{c}$=9.94\,MHz which is generated from a 480-nm diode laser (Toptica, DL pro). The coupling laser is resonant with the atomic transition to maximize the read-out signal. The gain of the photodetector is set to different gain levels $10^{5}\,\mathrm{V/A}$ and $10^{6}\,\mathrm{V/A}$ resulting in a different system noise base of Fig.~\ref{fig2}c and the Fig.~\ref{fig3}.

To measure the properties of a single module of RF-chip-integrated Rydberg atomic receiver, for example, sensitivity, dynamic range, and instantaneous bandwidth. A strong LO RF field and a weak signal RF field (about -35\,dBm at the signal source) are first combined by a power divider and then fed into the metawaveguide chip through a Sub-Miniature A (SMA) connector port. Then the SPP waves of these RF fields are encouraged in the center of the chip (over the spoof-SPP units) and coupled with the excited Rydberg atoms. The probe and coupling beams over the chips are aligned close to the surface of the chip and aligned with the propagation direction of the RF signal. The position for the laser beams in the y-axis direction in Fig.~5 is carefully adjusted to overlap with the central of the slot structure which has the most uniform and the strongest field distribution over the chip to reduce the inhomogeneous effect of the RF field distribution. The phase and intensity of the traveling wave over the chip quickly vary, so the electric field that is finally detected by the atoms is averaged. In our system, the beams are about 1 mm above the chip. 

The LO RF fields in multi-band receiving test (marked as LO1 and LO2) are emitted by two signal sources (Rohde \& Schwarz SGS100A and Ceyear 1465F) and then are applied through a pair of horn antennas (HD-70WCAS, HD-180SGAH20S) instead of directly injecting to the input port of the chip to reduce the crosstalk of the LO RF fields between the two receiving modules. The LO power being input into the horn antennas is set to about 20\,dBm for 300\,MHz as the frequency has exceeded the working bandwidth for the horn. A larger LO power would significantly broaden the EIT spectrum thus reducing the signal-noise ratio. In this process, two signal RF fields are synthesized by two (Rigol DSG3136B and Ceyear 1465F) microwave signal sources, respectively. 

\section{Declarations}
\subsection{Acknowledgements}This work is supported by the National Key R\&D Program of China (2022YFA1404002), the National Natural Science Foundation of China (Grant Nos. U20A20218, 61525504, 61722510, 61435011), the Major Science and Technology Projects in Anhui Province (Grant No. 202203a13010001), and the National Natural Science Foundation of China (Grant No. 11934013).

\subsection{Conflict of interest}The authors declare no competing interests.
\bibliography{ref}
\end{document}